# Origin and Stability of exomoon atmospheres
## Implications for habitability

**Helmut Lammer[1], Sonja-Charlotte Schiefer[2,3], Ines Juvan[1,3], Petra Odert[3], Nikolai V. Erkaev[4,5], Christof Weber[1], Kristina G. Kislyakova[1], Manuel Güdel[6], Gottfried Kirchengast[2], Arnold Hanslmeier[3]**

[1]Space Research Institute, Austrian Academy of Sciences, Graz, Austria, [2]Wegener Center for Climate and Global Change, Univ. Graz, Austria, [3]Institute of Physics, IGAM, Univ. Graz, Austria, [4]Institute for Computational Modeling, Russian Academy of Sciences, Krasnoyarsk, Russian Federation, [5]Siberian Federal University, [6]Institute of Astronomy, University of Vienna, Austria

### Abstract

We study the origin and escape of catastrophically outgassed volatiles ($H_2O$, $CO_2$) from exomoons with Earth-like densities and masses of $0.1 M_\oplus$, $0.5 M_\oplus$ and $1 M_\oplus$ orbiting an extra-solar gas giant inside the habitable zone of a young active solar-like star. We apply a radiation absorption and hydrodynamic upper atmosphere model to the three studied exomoon cases. We model the escape of hydrogen and dragged dissociation products O and C during the activity saturation phase of the young host star. Because the soft X-ray and EUV radiation of the young host star may be up to ~100 times higher compared to today's solar value during the first 100 Myr after the system's origin, an exomoon with a mass $< 0.25 M_\oplus$ located in the HZ may not be able to keep an atmosphere because of its low gravity. Depending on the spectral type and XUV activity evolution of the host star, exomoons with masses between ~$0.25 - 0.5 M_\oplus$ may evolve to Mars-like habitats. More massive bodies with masses $>0.5\ M_\oplus$, however, may evolve to habitats that are a mixture of Mars-like and Earth-analogue habitats, so that life may originate and evolve at the exomoon's surface.

## Introduction

As of November 2014, the *Extrasolar Planets Encyclopedia*[1] lists 1160 planetary systems, 471 multiple planet systems and 1849 planets. Since this number is growing continuously, also in the "super-Earth"-mass/size domain, the question if some of these exoplanets can be considered as potentially habitable is clearly of great interest. Because the majority of confirmed exoplanets and candidates orbiting in the habitable zone (HZ) of their host stars are either super-Earths or gas giants, discussions came up if possible moons of such planets, termed exomoons, could be habitable too (e.g., Williams et al. 1997; Kaltenegger 2000; 2010; Scharf 2006; Barnes and O'Brien 2002; Cassidy et al. 2009; Kipping et al. 2009; Porter and Grundy 2011; Awiphan and Kerins 2013; Heller and Zuluaga 2013). The question of the habitability of exomoons is closely linked to the nature of their potential atmospheres, the detection of atmospheric biosignatures (e.g., Kiang et al. 2007; Grenfell et al. 2010; Kaltenegger 2010; 2013; Rugheimer et al. 2013) and the general exomoon detection possibilities (e.g., Kipping 2009a; 2009b; Kipping et al. 2012 and references therein; Heller and Barnes 2014a).

---

[1]http://exoplanet.eu/catalog/

**Detection Methods of Exomoons**

According to Heller and Barnes (2013), the detection of large exomoons with masses larger than that of Mars should now be feasible, and the development of detection methods is progressing. Heller and Barnes (2014b) give four main reasons why habitable exomoons are objects of interest

i. Provided that such large satellites exist in extrasolar systems, their masses should be $\geq 0.25 M_{\oplus}$ (Kipping et al. 2009) to be detectable with current methods, i.e. that detected objects could be more than twice as massive as Mars ($\sim 0.1 M_{\oplus}$).

ii. Because most exomoons should be tidally locked to their host planets, they have shorter days compared to their stellar year, which represents a big advantage regarding their habitability compared to tidally locked exoplanets in the HZs of M dwarfs.

iii. It is expected that an exomoon orbiting a massive planet in its equatorial plane will experience seasons more likely than an equidistant single terrestrial-type exoplanet (Heller et al. 2011; Porter and Grundy 2011).

iv. Current statistics of exoplanet detections in the Kepler data[2] indicate that there may be more potential hosts of moons (such as warm Neptunes and gas giants) than Earth-analogs, thus it seems to be plausible that potentially habitable exomoons could be even more numerous than Earth-analogue exoplanets themselves.

Here we give a brief overview how exomoons can be detected in orbits around giant exoplanets. In recent years, several methods have been developed to detect exomoons in transit light curves of giant exoplanets. As discussed by Kipping et al. (2011; 2012) there exist two broad categories of observational effects induced by an exomoon in a transiting planet system, namely dynamical variations and eclipse features. Another possibility is related to exoplanetary radio emissions that are modulated by exomoons.

---

[2] Planetary Habitability Laboratory: http://phl.upr.edu

*Transit Timing Variation*

The so-called transit timing variation (TTV) concept was first established by Sartoretti and Schneider (1999) and describes a phenomenon affecting the motion of the host planet around the barycenter of the planet-moon system. Consequently, if the planet is hosting an extrasolar moon, there exists the possibility that one can observe a variation in the transit timing of the orbiting planet. However, it should be noted that Trojan bodies (e.g., Dvorak 2004; Ford and Holman 2007; Schwarz et al. 2007), parallax effects (Scharf 2007), star spots (Alonso et al. 2008, Szabó et al. 2006), etc., can cause a signal similar to the presence of an exomoon. Because of the fact that the waveform of the TTV signal is undersampled, one cannot reliably determine the orbital period of the exomoon, but instead measures the rms amplitude of the TTV signal, which is proportional to both the satellite's mass and the planet-moon separation (Kipping et al. 2012). Therefore, as TTVs alone do not provide a unique solution, a second complementary method is needed which will be discussed below.

*Transit Duration Variation*

A second method complementary to TTVs is to look for transit duration variations (TDVs) which refer to the detection of a change in the transit duration of an exoplanet over many observations (Agol et al. 2005; Holman and Murry 2005). This is due to changes in the apparent velocity of the planet caused by the planet-moon interaction (Sartoretti and Schneider 1999). For systems with non-coplanar orbits, the TDV effect can be separated into two main components, namely in a velocity (TDV-V) and a transit impact parameter (TDV-TIP) component (Kipping 2009b). The TDV-V component is caused by the variation of the velocity of the planet due to the exomoon's gravity, whereas the TDV-TIP part, which is like the TTV an effect due to a change in position, can be explained by the fact that if the planet-moon orbital plane is not exactly normal to the sky, the planet's reflex motion leads to variation in the apparent transit impact parameter and TDVs will follow. This TDV-TIP component is smaller than the TDV-V signal, but it allows one to gain the direction of the orbital motion of the satellite (Kipping 2009b). Furthermore, Kipping (2011) describes the total TDV signal as a linear combination of TDV-V and TDV-TIP components and it should be noted that for a system with an orbital inclination of 90° and a circular orbit the TDV -TIP component will be zero. In addition, these signals, which have similar amplitudes, also show a 90°-phase difference which provides a unique exomoon signature and as a result, a combination of TDV and TTV can confirm the presence of an exomoon (Kipping 2011, Kipping et al., 2012).

Furthermore, it should be noted that the ratio of both TDV and TTV allows the determination of the exomoon's mass. Kipping et al. (2012; 2013) discussed Kepler's sensitivity of detecting to a gas giant in the HZ with a single exomoon, con-sidering exomoon masses $\geq 0.2 M_\oplus$. They found that an exomoon with $0.2 M_\oplus$ may be detectable if it orbits a Saturn-like exoplanet transiting an M dwarf. Moreover, they found that a similar exomoon orbiting a Jupiter-like exoplanet inside a G-star HZ could not be discovered in the Kepler data by the TTV-TDV method. These conclusions are also supported by Awiphan et al. (2013) who modeled light curves of systems comprising M-dwarf hosts of mass $0.5 M_{Sun}$ and radius $0.55 R_{Sun}$ with Jupiter-like giant planets hosting rocky "super-Earth"-exomoons. They found that satellites orbiting planets within the HZ of an M-dwarf can produce both detectable TTV and TDV signatures with Kepler-class photometry.

### Eclipse Features

A third possibility to detect exomoons is the search for eclipse features of the exomoon, which can be either in-front of the star (auxiliary transits) or in-front/behind the planet during the star-planet transit (mutual event) (Kipping et al. 2012). These eclipses are sensitive to the exomoon's radius which offers the opportunity to determine the density of the exomoon, if the mass of the host planet is known. In addition, it should be noted that eclipse features due to an exomoon can not only be detected in stellar light curves in general, but also in phase-folded transit light curves (Heller 2014) as well as in a planetary spectrum (Heller and Albrecht 2014). These authors showed that direct imaging of exomoons transiting in front of young self-luminous gas giants will be doable with the European Extreme Large Telescope (E-ELT) and time-resolved spectroscopy may even be enable the determination of an exomoons sense of orbital motion.

### Radio Emissions

In case extra-solar gas giants have large exomoons that act as a strong plasma source for the system, this opens the possibility for the detection of radio emission and therefore the identification of a strong magnetosphere of the host planet (Rucker 2001; Grießmeier et al. 2007; Hess and Zarka 2011). Such a magnetosphere on the other hand may then act as a protection of the exomoon surface against energetic particles (Heller and Zuluaga 2013). It is known that the Galilean satellite Io triggers the strongest radio emission at Jupiter with frequencies up to 40 MHz. This volcanic moon acts as a main plasma source for the Jovian magnetosphere and the plasma interaction enhances the radio signal.

Because a radio signal of an extra-solar gas giant with an exomoon that acts as a strong plasma source like Io would be enhanced, Nichols (2011; 2012) studied such configurations. It was found that the radio power for Jovian exoplanets with internal plasma sources increases with the rotation rate of the planet and with the soft X-ray and extreme ultraviolet (XUV) luminosity of the system's host star. For all of the studied XUV luminosities this author finds detection possibilities beyond one pc, where the planets should orbit at 1 to 50 AU. For XUV luminosities that are 100 - 1000 times higher than the solar value the detection threshold is about 20 pc and if the planet is rapidly rotating this might further extend up to 50 pc.

The detection of big exomoons via the modulation of the host exoplanet's radio emission could be possible with the Long Wavelength Array (LWA) and the Low Frequency Array (LOFAR) (Noyola et al. 2014). However, they find that such a body would have to be as large as ~3.5$R_\oplus$, i.e. approximately as large as Uranus. Thus, systems like this in which the smaller body could be detected by the modulation of the radio emission of the larger body are more like a double planet system. Although the results given in Noyola et al. (2014) are slightly optimistic one should keep in mind that such systems may be rare so that this technique is probably not the most efficient exomoon detection technique, at least in the near future.

## Aspects of Exomoon Habitability

Although convincing observational evidence of the existence of exomoons is still missing (e.g., Kipping et al. 2012), as mentioned before, many studies are not only focusing on exomoon detectability, but also on various habitability aspects concerning these potentially exciting objects. This comprises topics like

-   tidal heating and energy sources (e.g., Reynolds et al. 1987; Scharf 2006; Cassidy et al. 2009; Heller 2012; Heller and Barnes 2013; 2014b),
-   orbital stability (e.g., Barnes and O'Brien 2002; Donnison 2010a; 2010b; Weidner and Horne 2010; Cuntz et al. 2013; Forgan and Kipping 2013; Donnison 2014; Gong et al. 2014),
-   illumination effects (Heller and Barnes 2013; Forgan and Yotov 2014),
-   Hill stability (e.g., Heller and Barnes 2014b; Donnison 2010a; 2010b; Hinkel and Kane 2013),
-   magnetic environment (Heller and Zuluaga 2013),
-   formation scenarios (e.g., Kipping et al. 2012; Heller and Pudritz 2014),

and

-   biosignatures (Kiang et al. 2007; Grenfell et al. 2010; Kaltenegger 2010; 2013; Rugheimer et al. 2013).

According to Heller and Barnes (2013), one conclusion that can be drawn from these studies is that one has to distinguish between habitability aspects of exoplanets and exomoons. The main differences occur when talking about the definitions of their HZs as well as their energy sources (Heller and Barnes 2013).

## The Habitable Edge and Orbit Stability

Besides these distinctions, there also exists an analogy to the concept of the circumstellar HZ, which was described by Kasting et al. (1993), namely the so-called habitable edge (HE). The HE is defined as the minimum orbital separation between a planet and its moon to let the satellite be habitable in presence of stellar and planetary illumination and tidal heating. Consequently, this means that potentially habitable exomoons have to orbit their planetary hosts outside of the HE (Heller and Barnes 2013). Furthermore, the Roche limiting radius determines how close an exomoon can approach the planet before tidal disruption occurs, while the Hill stability of the star–planet–moon system determines stable orbits of the moon around the planet (Donnison 2010a; 2010b; 2014; Hinkel and Kane 2013). In addition, several studies (Kipping et al. 2012; Kipping 2011; Lews 2011) discuss the existence of two different classes of exomoons, namely regular satellites that formed, like Io, Europa, Ganymede and Callisto, together with their host planet (e.g., Canup and Ward 2002), or so-called irregular satellites that have either been captured or formed by giant impacts (i.e., Triton, Moon, etc.), with the distinction made regarding their origin and evolution.

The first search for hypothetically habitable exomoons with Kepler was recently initiated by Kipping et al. (2013a; 2013b). This quest tries to assess whether exomoons with masses $\geq 0.25 M_{\oplus}$ are common objects or not, considering that such large satellites do not exist in our Solar System. Heller and Barnes (2014b) state that it is not yet clear if moons that massive (Mars-mass or ten times the mass of Ganymede) are plausible, but considering the many unexpected detections of gas giants in extrasolar systems, they claim that existence of a Mars-sized moon around a Neptune-sized planet may not be implausible. Heller and Barnes (2013) suggest that adequate host planets do exist and that combinations of all currently available observational and theoretical techniques can, in principle, estimate an exomoon's potential for habitability.

An important requirement that a regular or irregular exomoon may evolve to a habitat is its long-time orbital stability (e.g., Weidner and Horne 2010; Cuntz et al. 2013; Forgan and Kipping 2011; Donnison 2014; Gong et al. 2014). Barnes and O'Brien (2002) showed that hypothetical exomoons with $1 M_{\oplus}$ around Jupiter-mass gas giants are dynamically stable for the lifetime of the Solar System in star systems where the stellar mass is greater than >0.15 $M_{\odot}$.

Weidner and Horne (2010) studied the orbital stability of hypothetical exomoons of 87 known transiting gas giants. It was found that for 92% of the sample, exomoons larger than Earth's Moon could be excluded on prograde orbits, unless the host planet's internal structure is very different from the Solar System gas giants. In this study only WASP-24b, OGLE2-TR-L9, CoRoT-3b and CoRoT-9b could provide stable orbits for exomoons with masses > $0.4 M_{\oplus}$.

For instance, Cuntz et al. (2013) investigated the possibility of stable orbits for habitable exomoons around the Jupiter-mass gas giant HD 23079b, which has a low-eccentricity orbit in the outer region of HD 23079's HZ. These authors found that for the minimum mass case of an exomoon, the orbital stability limits of the Jupiter-mass gas giant HD 23079b could vary between 0.05236 and 0.06955 AU for prograde orbits and between 0.1023 and 0.1190 AU for retrograde orbits, corresponding to 0.306 and 0.345 (prograde orbits) and 0.583 and 0.611 (retrograde orbits) of the exoplanet's Hill radius. Larger stability limits for exomoons were obtained if they assumed a higher mass for the Jupiter-type planet. The general findings can be summarized that the larger the semimajor axis, the larger the stability limit. It was also found that the larger the eccentricity, the smaller the stability limit. A small number of exceptions have been identified if the orbital motion of the exomoon was affected by resonances (Cuntz et al. 2013). The before mentioned orbital stability limits are related to the outer orbit, while the inner limit of orbital stability is given by the exoplanet's Roche limit (Donnison 2014).

To summarize, all exomoon studies so far focused on the before mentioned subjects relevant for habitability, but did not investigate the stability of exomoon atmospheres against the high X-ray and extreme ultraviolet radiation (XUV) of the young host star. Because moons form either simultaneously with their planet hosts, are captured, or created by collisions during the host planet formation, their protoatmospheres will be exposed to XUV radiation that can be up to ~100 times higher compared to the XUV flux that Earth obtains from the present-day Sun (e.g., Ribas et al. 2005; Güdel 2007; Claire et al. 2012), similar to a planet's atmosphere. Depending on the host star's XUV evolution, the mass and size of the moon, as well as its atmospheric composition, this high-activity phase of the host star could lead to rapid escape of the exomoon's atmosphere so that its habitability conditions could change dramatically. On the other hand, if the moon is too massive it may not lose its primordial atmosphere, so that its habitability conditions will also be affected.

For example, if Saturn together with its large moon Titan migrates from its present orbital location at 10 AU to that of the Earth in 1 AU, Titan's nitrogen atmosphere would be exposed to an XUV flux that is ~100 times stronger compared to that of its present location. During the early stages of the Solar System, such a moon would have been exposed to an XUV flux which was up to ~$10^4$ times higher compared to its present location. The consequences of this high XUV emission onto a Titan-like body would be tremendous. As it was shown in Lichtenegger et al. (2010) and Lammer et al. (2011) a nitrogen atmosphere would not even be stable on an Earth-type planet that is exposed to such high XUV fluxes, indicating that the early atmosphere of the Earth was not nitrogen-dominated. A Titan-like exomoon in an orbit of a migrating gas giant that moves into an orbit location of 1 AU around a Sun-like star would therefore be transformed very fast due to efficient atmospheric escape processes into a dry, rocky body similar to Earth's moon.

Large exomoons, as suggested by Kipping et al. (2009; 2012), Kaltenegger (2010); Heller and Barnes (2014a; 2014b) and others, will likely have a hydrogen-dominated protoatmosphere (e.g., Ikoma and Genda 2012; Lammer et al. 2014), either nebula-based (similar to its host planet, like a warm Neptune or gas giant), or by outgassing of volatiles, such as $H_2O$, $CO_2$ and $CH_4$, during the solidification of its magma ocean (e.g., Elkins-Tanton 2008; Erkaev et al. 2013; 2014).

The aim of this work is the first attempt to take a closer look into this important aspect that has not been studied so far in detail. We will focus on potential exomoons orbiting large host planets inside the HZ of a young Sun-like G star. In our study we assume that the orbiting exomoons fulfill dynamical stability criteria so that they have stable orbits beyond the above mentioned HE and inside the HZ of their host planet, that is assumed to be a Jovian-type gas giant. In this study we focus on the most extreme phase of an exomoon's atmosphere evolution, but do not follow its evolutionary track after the young host star's XUV saturation phase ends and the radiation decreases to lower values. We also do not address possible plasma induced non-thermal atmospheric escape processes, as well as escape caused by photochemically produced suprathermal atoms (e.g. Lammer 2013 and references therein). As discussed in Lammer (2013), XUV-powered hydrodynamic escape driven by the young star is the most efficient atmospheric loss process of initial atmospheres. Therefore, we focus only on this loss process in this work. First, the XUV environment of young Sun-like stars is briefly discussed. Then we address the expected outgassing of volatiles into a young exomoon's atmosphere. After that we apply an upper atmosphere XUV-absorption and atmospheric escape model taking into account the enhanced XUV emission of a young solar-like star and estimate the escape of the atmospheres from hypothetical exomoons with masses of ~$0.1 M_\oplus$, ~$0.5 M_\oplus$ and ~$1 M_\oplus$.

## The X-ray and EUV environment of young Sun-like stars

All of the previous exomoon habitability studies neglected the high stellar radiation in the short wavelength range of young stars. Atmospheric escape is mainly driven by the XUV emission of the young host star (Lammer 2013). The evolution of this high-energy emission of young solar-like stars is divided into two regimes,

    i.    the saturation phase and the
    ii.   post-saturation evolution.

During the so-called saturation phase the stellar X-ray flux does not scale with the stellar rotation period and is saturated at ~0.1% of the bolometric luminosity $L_{bol}$ (Pizzolato et al. 2003; Jackson et al. 2012). During the post-saturation phase, the stars spin down from short periods of ~2 days due to angular momentum loss by the stellar wind (e.g., Güdel 2007). During this phase, the XUV emission of Sun-like stars is determined by their rotation periods. The evolution of the high-energy radiation of Sun-like stars has been studied in great detail using multi-wavelength observations of solar proxies with ages of ≥0.1 Gyr (e.g., Ribas et al. 2005; Güdel 2007; Claire et al. 2012).

    During the activity saturation phase of a young star, the XUV flux evolution is, as briefly mentioned before, mainly determined by $L_{bol}$. Between the zero-age main sequence, typically reached at about 50 Myr by a Sun-like star (Baraffe et al. 1998; Siess et al. 2000), and the end of the saturation phase, the stellar XUV flux is roughly constant at a value about 100 times the current solar value (Güdel 2007; Erkaev et al. 2014). Because gaseous planets form and migrate within a time peri-

od of a few Myr (e.g., Alibert et al. 2010), one can assume that hypothetical exo-moons around exosolar gas giants should also be formed during such short time scales (see also Heller and Pudritz 2014).

To better understand if large exomoons can protect their initial water inventory from high-energy stellar irradiation, we study the atmospheric escape of cata-strophically outgassed volatiles from solidifying magma oceans that are exposed to an XUV flux of the host star that is ~100 times higher compared to that of to-day's solar value at Earth.

## Volatile Outgassing During the Solidification of Magma Oceans

The four large Galilean satellites, Io, Europa, Ganymede and Callisto are a by-product of the formation of Jupiter (e.g., Polack and Reynolds 1974; Canup and Ward 2006; Lewis 2011). Jupiter formed during the lifetime of the gaseous solar nebula, which was of the order of ~1–10 Myr (e.g., Haisch et al. 2001; Montmerle et al. 2006). The internal structures of these large moons are comparable to those of the terrestrial planets (e.g., Schubert et al. 2004). The silicate and metal materi-als that make up planetary bodies including large moons was most likely melted once or several times during their formation process (e.g., Albarède and Blichert-Toft 2007; Elkins-Tanton 2012).

Melting can be caused by radiogenic heating from short-lived radioisotopes, gravitational heating of the accreted material, and the energetics of accretionary impacts (e.g., Urey 1955, Safronov 1969; Wetherill 1980; LaTourrette and Was-serburg 1998; Halliday et al. 2001; Albarède and Blichert-Toft 2007; Elkins-Tanton 2008; 2011; 2012). As a consequence, the formation of magma oceans creates also compositional differentiation that affects the final volatile contents of the planetary building blocks, including large planetesimals, planetary embryos, protoplanets and large moons, whose radii range from tens to hundreds, or even thousands of kilometres. In the Solar System the process of planetary differentia-tion has occurred on all planets, dwarf planets, the asteroid Vesta, and main natu-ral satellites, such as Earth's Moon. The Galilean satellites are all considered to be differentiated, although Callisto only partly so (e.g., Canup and Ward 2002; Schu-bert et al. 2004).

The partial or complete melting related to planetary differentiation is the pro-cess of separating out different constituents of a planetary body as a consequence of their physical or chemical behaviour, where the body develops compositionally distinct layers; the denser materials of a planet sink to the centre, while less dense materials rise to the surface. Such a process tends to create a core and mantle in the planetary body. Thus, from the knowledge gained from Solar System bodies, we can assume that magma ocean formation will also occur at the earliest evolu-tionary stages of large moons orbiting exoplanets.

Volatiles such as $H_2O$ and $CO_2$ are integrated in magma ocean liquids and as solidification proceeds they are degassed into a growing steam atmosphere. Ohtani et al. (2004) and Wyllie and Ryabchikov (2000) found that at pressures and temperatures of magma ocean crystallization no hydrous or carbonate minerals will crystalize. The quantity of $H_2O$ and carbon compounds (e.g., $CO_2$, $CH_4$, CO) available for degassing depends on the bulk composition of the magma ocean. The solidification of a magma ocean starts at the bottom because the steep slope of the adiabat with respect to the solidus in pressure-temperature space causes them to intersect first at depth (e.g. Walker et al. 1975; Solomatov 2000; Elkins-Tanton 2008; 2011; 2012). The larger the pressure ranges of a planetary body, the stronger the effect.

The initial amount of $H_2O$ and carbon species delivered during the formation of our hypothetical exomoons is not well constrained. For this reason we assume initial compositions as expected for Mars and Earth (Jarosewich 1990; Elkins-Tanton 2008; Brasser 2013). According to Brasser (2013), a planetary embryo, such as Mars in its present orbit, may have had an initial composition of $\sim 1000 - 2000$ ppm $H_2O$, with about 1/5 carbon content. An embryo of similar size which originates at a closer orbital distance would have most likely lower water content. Elkins-Tanton (2008) showed that for a range of magma ocean bulk compositions of $\sim 500$–5000 ppm $H_2O$, $\sim 70$–99 % of the initial $H_2O$ and carbon content is degassed into the planetary surroundings.

Elkins-Tanton (2008) and Erkaev et al. (2013) used initial compositions as discussed above and modelled expected steam atmospheres that are produced during the solidification of magma oceans with depths of $500 - 2000$ km.

**Table 1:** Minimal and maximal expected atmospheric partial surface pressures $P_{H2O}$ and $P_{CO2}$ in bar of catastrophically outgassed steam atmospheres for initial $H_2O$ ($0.05 - 0.1$ wt%) and carbon contents ($0.01 - 0.02$ wt%) of magma oceans with depths of 500 - 2000 km (Elkins-Tanton 2008; Erkaev et al. 2013; Erkaev et al. 2014), estimated for exomoons with $0.1 M_\oplus$, $0.5 M_\oplus$ and $1 M_\oplus$ located within the HZ of a solar-like star at $\sim 1$ AU.

| Exomoon | $P_{H2O}$ [bar] | $P_{CO2}$ [bar] |
|---|---|---|
| $0.1 M_\oplus$ \| $0.46 R_\oplus$ | $\sim 30 - 120$ | $\sim 7 - 25$ |
| $0.5\ M_\oplus$ \| $0.8 R_\oplus$ | $\sim 50 - 300$ | $\sim 10 - 65$ |
| $1.0 M_\oplus$ \| $1 R_\oplus$ | $\sim 75 - 460$ | $\sim 35 - 100$ |

Table 1 shows partial surface pressure ($P_s$) ranges of catastrophically outgassed steam atmospheres for the three test-exomoons with expected initial $H_2O$ content of $\sim 0.05 - 0.1$ wt% and carbon content of $\sim 0.01 - 0.02$ wt%, with assumed bulk magma ocean depths of $\sim 500 - 2000$ km. These values consider materials from dry to wet and are in agreement with the modelled values of Elkins-Tanton (2008) and Erkaev et al. (2013; 2014).

In the following section we will estimate the critical size and mass of an exomoon inside the HZ above which outgassed volatiles (e.g. $H_2O$, $CO_2$) can be maintained during the early stages characterized by high activity of the host star. We study three hypothetical exomoons with an average Earth-like density of 5.5 g cm$^{-3}$, masses $M_{em}$ of $0.1M_\oplus$, $0.5M_\oplus$ and $1M_\oplus$ and corresponding radii $R_{em}$ of $0.46R_\oplus$, $0.8R_\oplus$ and $1R_\oplus$.

## Volatile Loss from Exomoons in the Habitable Zone

After the phase of catastrophic volatile outgassing of an exomoon, high levels of the young star's short-wavelength radiation will lead to dissociation of the $H_2O$ and $CO_2$ molecules (Chassefiere 1996a; Tian et al. 2008; Lammer 2013). Therefore, one can expect that the young exomoon has a hydrogen-dominated upper atmosphere. Such an atmosphere will expand hydrodynamically and escape efficiently (e.g., Watson et al. 1981; Chassefière 1996a; 1996b; Tian et al. 2005; Lammer et al. 2013; 2014; Erkaev et al. 2013; 2014).

### Model, Boundary Conditions and Hydrogen Escape Rates

For studying the response of the hydrogen-dominated upper atmospheres of the three hypothetical exomoons to an XUV flux that is about 100 times higher compared to that of the Sun in 1 AU, we apply a time-dependent 1-D hydrodynamic upper atmosphere model that solves the system of fluid equations for

i.   mass,
ii.  momentum,
iii. and energy conservation in spherical coordinates (e.g., Erkaev et al. 2013; 2014; Lammer et al. 2013; 2014).

We also take into account the possible reduction of the gravitational potential caused by the presence of the massive host planet. The gravitational acceleration

$$g = -\nabla \phi. \qquad (1)$$

is derived from the gravitational potential $\phi$

$$\phi = -G\frac{M_{em}}{r} - G\frac{M_{pl}}{(d-r)} - G\frac{M_{pl}+M_{em}}{2d^3}\left(\frac{M_{pl}\,d}{M_{pl}+M_{em}} - r\right)^2, \qquad (2)$$

where $M_{em}$ is the mass of the exomoon, $M_{pl}$ the mass of the host planet, $d$ the orbital distance between the exomoon and its host planet, and $r$ the radial distance from the moon's center. The numerical scheme used for solving the system of equations, as well as the applied energy absorption model for calculating the XUV volume heating rate, are described in detail in Erkaev et al. (2013; 2014) and Lammer et al. (2013; 2014). In our energy absorption model we adopt the same fraction of absorbed XUV radiation that is transformed into thermal energy as Chassefière (1996a), who studied the escape of water from early Venus and found that the so-called XUV heating efficiency is most likely close to ~15 %.

The lower boundary conditions of the simulation domain are set by fixing the gas temperature and the number density at the base of the thermosphere $r_0$. The number density at $r_0$ is strictly determined by the XUV absorption optical depth of the upper atmosphere and is in the order of $5 \times 10^{12}$ cm$^{-3}$ (Tian et al. 2005; Erkaev et al. 2013; 2014). Because of the high XUV flux H$_2$ molecules will be dissociated close to the lower boundary level at an atmospheric pressure of ~ 1 μbar. Therefore, the majority of the upper atmosphere will be populated by atomic hydrogen. The temperature is more or less similar to the equilibrium temperature of ~250 K at the exomoon/host planet location in the HZ at 1 AU.

**Table 2:** Modeled hydrodynamic escape rates of atomic hydrogen, $L_H$, from exomoons with masses $M_{em}$ of $0.1M_\oplus$, $0.5M_\oplus$ and $1M_\oplus$ and radii $R_{em}$ of $0.46R_\oplus$, $0.8R_\oplus$ and $1R_\oplus$ located within the HZ of a solar-like star at 1 AU.

| Exomoon $M_{em}$ \| $R_{em}$ | H escape rate $L_H$ [s$^{-1}$] |
|---|---|
| $0.1M_\oplus$ \| $0.46R_\oplus$ | ~$2.0 \times 10^{33}$ |
| $0.5 M_\oplus$ \| $0.8R_\oplus$ | ~$1.0 \times 10^{32}$ |
| $1.0M_\oplus$ \| $1R_\oplus$ | ~$4.5 \times 10^{31}$ |

Because of the hot environment of the catastrophically outgassed steam atmosphere above the molten or partly molten magma ocean, the base of the thermosphere rises to an altitude $z_0 = r_0 - R_{em}$ where the atmospheric pressure or density has a similar value as for a cooler surface environment. Depending on the exomoon's surface gravity, the altitude $z_0$ above a hot molten surface is expected to be located at 100 - 1000 km above the hot surface (Marcq 2012; Erkaev et al. 2014).

Because the coupled magma-ocean–steam atmosphere modeling of Marcq (2012) depends on some uncertain input parameters, we model the escape rates for both $z_0 = 100$ km and 1000 km and use the average of the resulting escape rates for the atmospheric loss study. The modeled H escape rates for degassed steam atmospheres that are exposed to a 100 times higher XUV flux than that of the present solar value at 1 AU are shown in Table 2 for the three studied exomoons.

One can see from Table 2 that the H escape rates reaches an extremely high value of $>10^{33}$ s$^{-1}$ for the Mars-like exomoon, about an order of magnitude lower for a mass of $0.5M_\oplus$, and ~$4.5 \times 10^{31}$ s$^{-1}$ for the Earth-like body. We note that the escape rates may decrease when the exomoon moves behind the host planet so that its atmosphere is temporarily protected from the star's irradiation. However, the expected decrease of the escape rate is small compared to the other uncertainties, such as the intrinsic variability of the stellar XUV flux and particle processes that may also contribute to the heating of the exomoon's thermosphere (e.g., energetic particle interaction processes related to the host planet's radiation belts, etc.). Furthermore, depending on the orbit location where the exomoon originates its average density could also be lower compared to the 5.5 g cm$^{-3}$ assumed in our study. However, exomoons with lower average densities between $2 - 3$ g cm$^{-3}$ would experience higher hydrogen loss rates than that shown in Table 2. On the other hand such bodies will be more volatile-rich, but their larger reservoir is most likely also be lost very fast due a more efficient escape rate. To summarize, we expect that the atomic hydrogen escape rates shown in Table 2 should be considered as average values with accuracy to within a factor of 2.

Because the modeled hydrogen escape rates depend on the atmospheric composition one cannot directly compare the loss rates of steam atmospheres with that of present Venus, Earth or Mars. First, because of the much weaker solar XUV flux the present planetary atmospheres do not lose their hydrogen by hydrodynamic escape, but by the weaker Jeans-type loss (Bauer and Lammer 2004). On present Earth, hydrogen escape is also limited by diffusion through the bulk atmosphere up to the homopause and yields a value of $\leq 10^{27}$ s$^{-1}$ (Bauer and Lammer 2004). One can see that present Earth's H escape rate is ∽45 000 times lower than the escape rate from a hydrogen dominated upper atmosphere that is exposed to a 100 times higher XUV flux, as shown in Table 2. This escape rate would be similar for a hydrogen-dominated early Earth or an Earth-like exomoon orbiting a Jovian-type exoplanet at ∽1 AU inside the HZ of a G star. The H escape rate from present Venus' dry atmosphere, which obtains twice as much XUV compared to Earth, is ∽4 ×10$^{25}$ s$^{-1}$ (Lammer et al. 2006), and Mars loses H with a rate of ∽1.5 ×10$^{26}$ s$^{-1}$ due to its lower gravity (Lammer et al. 2003a).

We can also compare the escape rate of the Mars-type exomoon (Table 2) with the escape rate of a steam atmosphere that was most likely outgassed on early Mars, where the expected XUV flux of the young Sun at 1.5 AU was about 44 times higher compared to that of today's Sun at 1 AU. According to Erkaev et al. (2014), by considering similar initial atmospheric conditions, the H escape rate from a hydrogen-dominated Martian protoatmosphere at 1.5 AU would have been ∽7 × 10$^{32}$ s$^{-1}$, ∽2.8 times lower compared to a similar body at 1 AU.

If we assume that the exomoons orbit a Jupiter-like gas giant at a Callisto-like orbit at ∼ 25$R_{\mathrm{Jup}}$ or beyond, then the first Lagrangian point $L_1$ is located between $16.5 - 17R_{\mathrm{em}}$. This distance is rather small, but well beyond the distance in the upper atmosphere where the gas reaches the sonic point, so that Roche lobe overflow does not dramatically influence the escape rates in the studied scenarios. Because for other orbits and different masses of both the host planet and/or the exomoon, the Roche lobe may affect the escape rates more significantly and we plan to study this effect in more detail in the future.

However, as shown in Erkaev et al. (2007), parameter configurations with pronounced Roche lobe effects enhance the escape rates. In such cases the hydrogen loss rates would be higher than for those scenarios shown in Table 2. One can imagine that in some system configurations where an exomoon has a stable orbit near the HE, and thus closer to its host planet than in the present study, the exomoon may deliver its atmosphere via Roche lobe overflow into the atmosphere of its host planet. This atmospheric influx may modify the minor chemical composition of the exoplanet's atmosphere. In the following subsection we will use the atomic H escape rates to study the expected dragging and escape of heavier dissociation products, such as O and C atoms.

**Dragging and Escape of Heavy Dissociation Products by Atomic Hydrogen**

In an outgassed hot steam atmosphere, the $H_2O$ and $CO_2$ molecules in the upper atmosphere will be dissociated by the high UV and XUV fluxes of the young star and by the high temperature in the lower thermosphere caused by frequent impacts (e.g., Chassefière 1996a; Tian et al. 2005; 2009; Erkaev et al. 2013; 2014; Lammer 2013). Because of these processes O and C atoms will be incorporated in the expanding, upward-flowing hydrogen so that they may escape together.

Another important factor that is connected to the dragging efficiency of heavier dissociation products is the timescale during which the degassed volatiles remain in steam form. Lebrun et al. (2013) modeled the thermal evolution of magma oceans and estimated the cooling timescales of outgassed steam atmospheres for Venus, Earth and Mars. They concluded that water vapor condenses and may form lakes and even an ocean after ~1, ~1.5 and ~10 Myr for Mars, Earth, and Venus, respectively. For an Earth-like planet located at ≤0.66 AU from the Sun, the volatiles would remain vaporous forever.

However, these results are model-dependent, and other studies (e.g. Hamano et al. 2013) obtain longer timescales for an Earth-like planet at 1 AU (~4 Myr or even longer), in particular if frequent impacts during the first 100 Myr after the formation of the system are taken into account. Frequent impacts of large planetesimals, which were neglected in Lebrun et al. (2013), would most likely extend the cooling timescales found in this study. Tidal heating can be an important additional energy source for satellites (Reynolds et al. 1987; Scharf 2006; Cassidy et al. 2009), in contrast to single planetary bodies in the HZ. In the case of exomoons, tidal heating may keep their surface temperatures at a higher level for a longer time compared to usual planetary bodies (Heller 2012; Heller and Barnes 2013; 2014**b**). Tidal heating depends on the moon's eccentricity, its size, and the semi-major axis of the moon's orbit around its host planet (Heller et al. 2011; Leconte et al. 2010; Ferraz-Mello et al. 2008). Depending on the particular tidal heating rate and considering existing uncertainties related to the coupled magma ocean-steam atmosphere models, one should keep in mind that the estimated steam atmosphere cooling time scales of ~1.5 − 4 Myr, as derived for single planetary bodies inside the HZ, may possibly last longer in the case of large exomoons.

Considering the partial surface pressures of $H_2O$ and $CO_2$ shown in Table 1 together with the hydrogen escape rates given in Table 2, and by assuming that the molecules are dissociated, we can calculate the amounts of atomic H, O, and C and determine their evolution over time. For the outgassed atmospheres one can define a mixing ratio $f_{O,C} = N_{O,C}/N_H$ (Zahnle et al.1990), where $N_H$ is the atmospheric inventory of the primary escaping species, that is atomic hydrogen in our case, while subscript "O,C" corresponds to the other constituents, O and C. The fractionation factor $x_{O,C} = L_{O,C}/(L_H f_{O,C})$ for an escaping atmosphere composed of two major (H,O) and one minor (C) species can then be described, according to Zahnle et al. (1990), by the following two equations. The fractionation factor of the heavy major species (here, O) is given by

$$x_O = 1 - \frac{GM_{em}(m_O - m_H)b_{H,O}}{\phi_H r_0^2 kT\,(1 + f_O)} = 1 - \frac{g\,(m_O - m_H)\,b_{H,O}}{\phi_H kT\,(1 + f_O)}, \tag{3}$$

where $M_{em}$ is the exomoon's mass, $m_{O,H}$ the atomic masses of O and H, $b_{H,O}$ the binary diffusion parameter of O in H, $\phi_H$ the hydrogen escape flux at the exomoon's surface in cm$^{-2}$ s$^{-1}$, $r_0$ the exomoon's radius, $T$ the mean atmospheric temperature, $k$ the Boltzmann constant, $G$ is Newton's gravitational constant, and $g$ the gravitational acceleration. The fractionation factor of a minor constituent (in our case, C) in the presence of two major constituents (H and O) can then be written as

$$x_C = \frac{1 - \dfrac{GM_{em}(m_C - m_H)b_{H,C}}{\phi_H r_0^2 kT} + \dfrac{b_{H,C}}{b_{H,O}}(1 - x_O)f_O + \dfrac{b_{H,C}}{b_{O,C}}x_O f_O}{1 + \left(\dfrac{b_{H,C}}{b_{O,C}}\right)f_O}. \tag{4}$$

From left to right, the four terms can be identified with the hydrogen drag, gravity, buoyancy versus the heavy major constituent, and drag by the heavy major constituent (Zahnle et al. 1990). By using the definition of the fractionation factors $x_O$ and $x_C$ we can calculate $L_O$ and $L_C$, the escape rates of O and C. Thus, the escape rates of the dragged O and C atoms can then be written as (Erkaev et al. 2014)

$$L_O = L_H f_O x_O = L_H f_O \left(1 - \frac{\mu_O - 1}{\mu_O \Phi_O\,(1 + f_O)}\right), \tag{5}$$

and

$$L_C = L_H f_C \frac{1 - \dfrac{\mu_C - 1}{\mu_C \Phi_C} + \dfrac{b_{H,C}}{b_{O,C}}f_O x_O + \dfrac{b_{H,C}}{b_{H,O}}f_O(1 - x_O)}{1 + \dfrac{b_{H,C}}{b_{O,C}}f_O}, \tag{6}$$

with $\mu_{O,C} = m_{O,C}/m_H$, and the parameter

$$\Phi_{O,C} = \frac{L_H kT}{3\pi GM_{em}\,m_{O,C}\,b_{H,(O,C)}}, \tag{7}$$

the ratio of drag to gravity, must be $> (\mu_{O,C}-1)/\mu_{O,C}$ for drag to occur. The factor $3\pi$ stems from the adopted solid angle over which we assume that the escape takes place (Erkaev et al. 2014). The binary diffusion parameter of O in H, $b_{H,O} = 4.8$ $\times 10^{17} T^{0.75}$ cm$^{-1}$s$^{-1}$, is taken from Zahnle and Kasting (1986). The binary diffusion parameter of C in H, $b_{H,C}$, is assumed to be equal to $b_{H,O}$. Due to lack of data, we assume that roughly $b_{O,C} = 2 \times 10^{17} T^{0.75}$ cm$^{-1}$s$^{-1}$. The evolution of the outgassed inventories of O and C, and hence their partial surface pressures, are found by numerically integrating Eqs. (5) and (6) over time.

Fig. 1 shows the temporal evolution of the partial surface pressures $P_{surf}$ of H, O and C normalized to the total initial surface pressure $P_{total} = P_{H2O}+P_{CO2}$ for the outgassed volatile amounts taken from Table 1 for the Mars-like exomoon (Fig. 1a: minimum values; Fig. 1b: maximum values). The more massive and larger exomoon with $0.5M_\oplus$ and $0.8R_\oplus$ is shown in Fig. 2 (Fig. 2a: minimum values; Fig. 2b: maximum values).

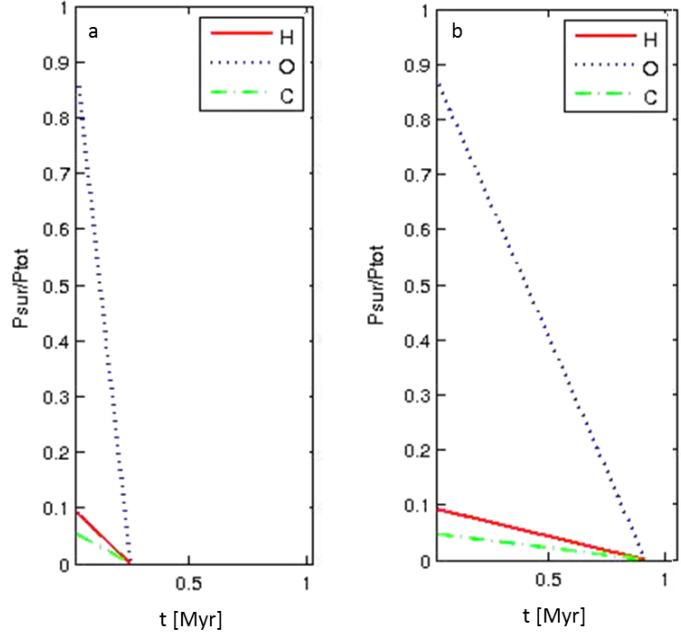

**Fig. 1:** Temporal evolution of the partial surface pressures $P_{surf}$ of H, O and C normalized to the total initial surface pressure $P_{total}$ for the expected outgassed volatile amounts of the Mars-like exomoon shown in Table 2. Panel a) corresponds to the lower degassed values, and panel b) to higher degassed volatile amounts. H$_2$O condensation of an outgassed steam atmosphere will, according to Lebrun et al. (2013) and Hamano et al. (2013), occur at ~1-4 Myr at ~1 AU.

As discussed above, a crucial factor is the timescale during which the outgassed steam atmosphere remains in steam form and after which time the water vapor begins to condense and rain out of the atmosphere so that lakes and even oceans are formed (Lebrun et al. 2013; Hamano et al. 2013). According to Lebrun et al. (2013) and Hamano et al. (2013) a magma ocean-related outgassed steam atmosphere will reach the condensation phase after ∽1-4 Myr if the object orbits a solar-like host star at ∽1 AU.

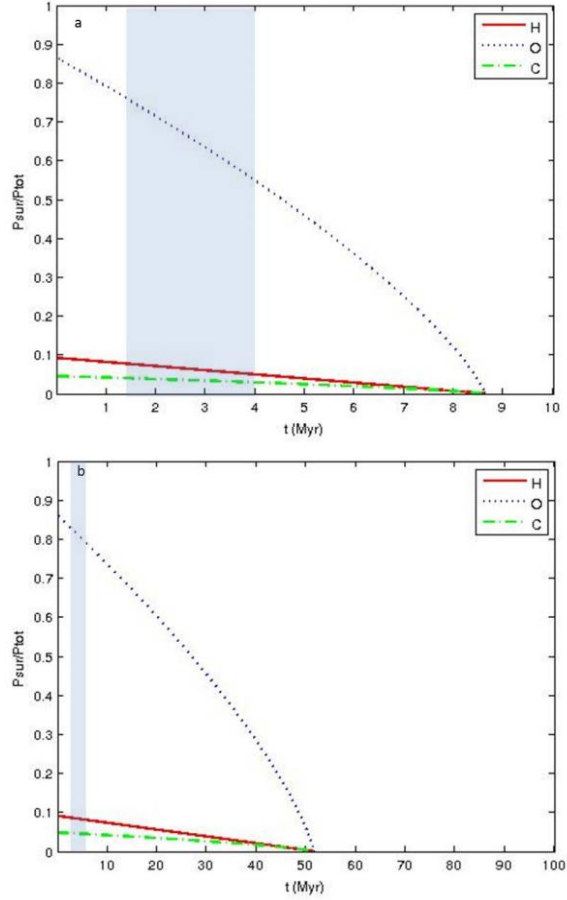

**Fig. 2:** Temporal evolution of the partial surface pressures $P_{surf}$ of H, O and C normalized to the total initial surface pressure $P_{total}$ for the expected outgassed volatile amounts shown in Table 2. Panels a) and b) correspond to the expected minimum and maximum outgassing scenarios for the exomoon with $0.5M_{\oplus}$ and $0.8R_{\oplus}$. The blue shaded area marks the range of water condensation onsets as modeled by Lebrun et al. (2013) (lower value: ~1 Myr) and Hamano et al. (2013) (higher value: ~4 Myr).

From our results shown in Figs. 1a and 1b one can see that Mars-like exomoons will lose their initial atmospheres before the steam atmospheres reach their condensation phase. Under such conditions, even an initial $H_2O$ inventory of ~ 120 bar will be lost in ≤1 Myr, while the condensation of a steam atmosphere at 1 AU may start ~1–4 Myr after its origin ($t$=0). The blue shaded area in Figs. 2a and 2b marks the time when $H_2O$, according to the before mentioned magma ocean steam atmosphere models, begins to condense because the outgassed steam atmosphere has cooled to the triple point of water (see also Lammer et al. 2011; Hamano et al. 2013; Lammer 2013; Lebrun et al. 2013).

One can see from Figs. 2a and 2b that hydrodynamic escape is less efficient for more massive exomoons with masses ≥$0.5M_\oplus$ compared to the Mars-like body. Depending on the initially outgassed volatiles, as well as additional amounts delivered later by impacts, exomoons with ~$0.5M_\oplus$ may lose the majority of their outgassed volatiles during the host star's saturation phase (≤0.1 Gyr), but due to their higher gravity the hydrodynamic escape rates are low enough so that formation of large lakes or oceans is also possible.

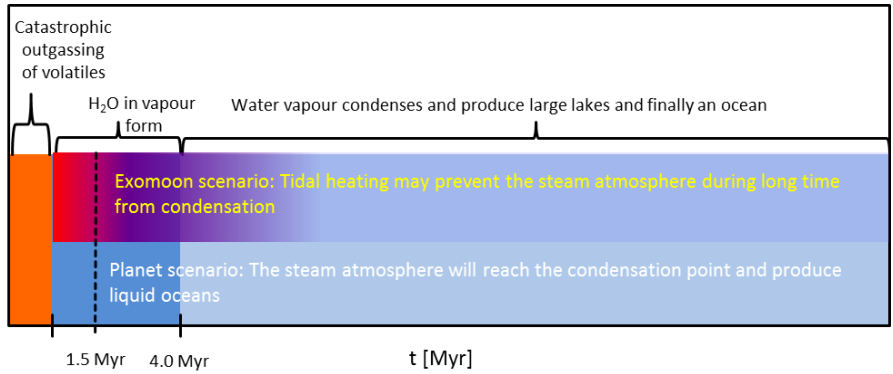

**Fig. 3:** Illustration of the timescales for escape and ocean formation on an Earth-like exomoon with $1M_\oplus$ orbiting a giant exoplanet inside the HZ of a young solar-like host star. Such a body will only lose a smaller fraction of the outgassed volatile contents shown in Table 1, because the time during which the steam atmosphere remains in steam form is short enough so that water will condense and form an ocean. Depending on orbital and planetary/exomoon parameters, on some exomoons tidal heating may keep the surface temperature hot enough so that the steam atmosphere may remain in vapour form for a longer time (several 10 – 100 Myr, or even for ever) compared to a single planet with similar mass and size. In such a case a large fraction or even the entire outgassed water inventory will escape to space.

However, one should note that the before mentioned tidal heating (Reynolds et al. 1987; Scharf 2006; Cassidy et al. 2009) may keep the surface of large exomoons hotter (Heller and Barnes 2014a; 2014b; Heller 2012) and therefore delay the onset of condensation of the steam atmosphere. Although a detailed study of this subject is beyond the scope of the present work one should consider that if tidal heating is efficient enough to prevent the steam atmosphere from condensation, a massive exomoon may evolve to a Venus-like body (see also Hamano et al. 2013).

We may expect that exomoons with masses of ~0.25 − 0.5$M_\oplus$ orbiting giant planets in the HZ most likely evolve to so-called Martian-like class II habitats (Lammer et al. 2009; Lammer 2013). For exomoons that reach the 1$M_\oplus$ domain the situation changes dramatically. Fig. 3 illustrates the scenario of the heaviest exomoon considered here which evolves more or less similar to the Earth-case shown in Fig. 3a of Lammer et al. (2011).

Thus, an Earth-like exomoon with 1$M_\oplus$ inside the HZ of a solar-like star will ultimately form oceans by outgassing and, after a ~1− 4 Myr period of strong atmospheric escape, by subsequent condensation and rainout (Lammer et al. 2011; Lebrun et al. 2013; Hamano et al. 2013), supplemented by further volatiles delivered by impacts (Albarède 2009) during later periods. However, possible tidal heating of the exomoon's surface will also alter the evolution, and hence the habitability, of Earth-like exomoons compared to a single Earth-like exoplanet at a similar orbital location in the HZ. In such a case one has to consider all the additional processes that have been addressed and studied in detail by Heller (2012) and Heller and Barnes (2014a; 2014b). We note that even if an Earth-like exomoon may form oceans, that does not necessarily mean that such a body would evolve further to an Earth-analogue class I habitat (Lammer et al. 2009; Lammer 2013). The geophysical and also climatological evolution of a tidally locked exomoon will certainly be different compared to a similar body that originated as an ordinary planet.

Moreover, the problem related to the stability of nitrogen atmospheres exposed to high XUV fluxes, as discussed in Lichtenegger et al. (2010) and Lammer et al. (2011), would also pertain to exomoons. One should also note that the results of this study would be different if the exomoons were orbiting a planet located in the HZ of more active K or M-dwarfs. M-dwarfs remain active in the XUV range longer than Sun-like stars and the whole planet-exomoon system would be exposed to denser stellar winds and coronal mass ejections (Scalo et al. 2007; Khodachenko et al. 2007; Lammer et al. 2007; 2009; Tian et al. 2009). Consequently, the atmospheric escape rates would be higher and remain so for a longer time period compared to a Sun-like star, so that the exomoon's atmosphere and surface environment would evolve differently.

From our results we suggest that in order to evolve to a habitat that lies between class II and class I (Lammer et al. 2009; Lammer 2013b), so that life may evolve on the surface, an exomoon should have a mass > 0.5$M_\oplus$. Objects with masses between ~ 0.25 to 0.5$M_\oplus$ may probably evolve to Mars-like class II habitats, while exomoons with lower masses will end most likely as large rocky moons with a thin gaseous layer that contains sputtered and desorbed surface material, similar to Earth's Moon or Mercury (e.g., Lammer et al. 2003; Killen et al. 2007; Wurz et al. 2012). Before we discuss our results in relation to the search for poten-tially habitable exomoons, one should also point out that both the outgassed volatile amounts and interiors of exomoons may be different from those studied here. Because our study is based on knowledge gathered in the Solar System, our results should be considered as a case study only.

## Habitability Aspects for Future Exomoon Discoveries

Recent studies investigated the detection limits of the above mentioned exomoon detection methods. Heller and Zuluaga (2013) discussed systems including Mars-like satellites ($\sim 0.1 M_\oplus$) orbiting Neptune-, Saturn- and Jupiter-sized host planets located in the center of the HZ of a K dwarf. They have chosen K-type instead of G-stars, because the latter are too bright and too massive to allow detection of an exomoon. These authors conclude that Kepler could be capable of detecting moons with masses of $\sim 0.2 M_\oplus$ by combining TTV and TDV signals.

Heller and Zuluaga (2013) also investigated the protection of an exomoon's atmosphere by the magnetosphere of the host planet, which may be relevant for its habitability. They concluded that exomoons as large as Mars, which are habitable in terms of illumination and tidal heating if they are orbiting a planet as large as Neptune, will hardly be affected by the planet's magnetosphere. Exomoons beyond an orbital location of $\sim 20 R_{pl}$ were found to be habitable in terms of illumination and tidal heating. Such exomoons will not be coated by the magnetosphere of their host planet within a timescale of $\sim 4.5$ Gyr. Hypothetical habitable exomoons with orbital separations between $\sim 5$ - $20 R_{pl}$ may be affected and possibly protected by the magnetosphere of a giant host planet.

However, as one can see from Figs. 1a and b, our results indicate that Mars-like exomoons or even bodies with a mass twice that of Mars, as assumed in the study of Heller and Zuluaga (2013), are most likely too small so that the majority of their magma ocean-related outgassed initial volatile content will escape to space during the active period of the young host star. Because the XUV flux remains high for a longer time for lower mass stars, even more massive exomoons inside the HZ of a K-star, as assumed by Heller and Zuluaga (2013), will experience problems related to the stability of their atmospheres. Due to XUV-heating and the resulting expansion of the upper atmosphere even a strong magnetosphere will not protect the exomoon's atmosphere against efficient atmospheric loss, because the majority of the gas that escapes is neutral. Subsequently it may be ionized and act as a plasma source in the system.

To conclude, exomoons may evolve to potential habitats with habitable surface conditions for life as we know it, if they have masses that are at least 20 times higher compared to the largest moon in the Solar System, Ganymede, with $\sim 0.025 M_\oplus$.

## Conclusions

We study the origin of catastrophically outgassed volatile-rich protoatmospheres ($H_2O$, $CO_2$) and their escape to space during the early stage after the formation of an exoplanet-exomoon system inside the habitable zone of a solar-like host star. Although tidal heating of the exomoon's surface is neglected our results show that exomoons with masses $\leq 0.1$ $M_\oplus$ lose all magma ocean-related outgassed volatiles before $H_2O$ condenses and forms liquid oceans on the moon's surface. The most important difference between a single sub- to Earth-like planetary body in the habitable zone and exomoons within the same size and mass range are possible Roche lobe effects that can enhance the atmospheric loss rate from the exomoon and more important the fact that the surface of a large exomoon, will stay longer or forever much hotter due to tidal heating. Because tidal heating is strongly dependent on the size of the moon and its orbital parameters one can also expect that this process will delay the onset of $H_2O$ condensation in the outgassed steam atmospheres for more massive exomoons. Besides these planetological issues the host stars evolution of the XUV flux is very relevant. We applied for our study an expected average XUV flux value of a young solar-like star that was initially 100 times (100 XUV) higher compared to that of today's Sun (Ribas et al. 2005). Because of this our atmosphere escape results should be considered as average G star scenarios. In case of initial XUV fluxes that are $< 100$ XUV the escape rates will be lower and for host stars where the XUV flux values are $> 100$ XUV the atmospheres would be lost much faster. Therefore, we conclude from our results that exomoons with masses of $> 0.1$ - $0.5 M_\oplus$ may evolve to Mars-like class II habitats and those with masses $> 0.5 M_\oplus$ to habitats that lie between class II and class I (Lammer et al. 2009; Lammer 2013).

**Acknowledgments:** H. Lammer, K. G. Kislyakova, S.-C. Schiefer, I. Juvan, N.V. Erkaev and M. Güdel, acknowledge the support by the FWF NFN project S116 "Pathways to Habitability: From Disks to Active Stars, Planets to Life". The same authors also acknowledge support by the related FWF NFN subprojects, S116 604-N16 "Radiation & Wind Evolution from T Tauri Phase to ZAMS and Beyond", S116607-N16 "Particle/Radiative Interactions with Upper Atmospheres of Planetary Bodies Under Extreme Stellar Conditions". N. V. Erkaev acknowledges support from the RFBR grant N 12-05-00152-a. H. Lammer, P. Odert and K. G. Kislyakova thank also the Helmholtz Alliance project "Planetary Evolution and Life." I. Juvan and C. Weber acknowledge also support from the European Astrobiology Association Network (EANA). P. Odert acknowledges support from the FWF-project P22950-N16. H. Lammer thanks L. Elkins-Tanton for discussions related to the formation of catastrophically outgassed volatiles and I. Juvan thanks R. Heller for fruitful discussions to exomoon studies in general. Finally, we acknowledge an unknown referee and R. Heller from the Department of Physics and Astronomy of the McMaster University in Hamilton, Ontario, Canada, for their suggestions and recommendations.